%% file: fpgaaas.tex
\documentclass[conference]{IEEEtran}
\IEEEoverridecommandlockouts
\usepackage[numbers,sort,compress]{natbib}

\usepackage[utf8]{inputenc} 
\usepackage[T1]{fontenc}    
\usepackage{hyperref}       
\usepackage{url}            
\usepackage{booktabs}       
\usepackage{amsfonts}       
\usepackage{nicefrac}       
\usepackage{microtype}      
\usepackage{lipsum}
\usepackage{graphicx}
\usepackage{subcaption}
\usepackage{xcolor}
\usepackage{multirow}
\graphicspath{ {./figures/} }
\usepackage{xspace}
\usepackage{amsmath}
\usepackage{doi}

\newcommand{\hlsfml}{\texttt{hls4ml}\xspace}
\newcommand{\python}{\textsc{Python}\xspace}
\newcommand{\grpc}{\texttt{gRPC}\xspace}
\newcommand{\unit}[1]{\ensuremath{\text{\,#1}}\xspace}
\newcommand{\faast}{FaaST\xspace}
\newcommand{\flarge}{\texttt{f1.x16large}\xspace}

\begin{document}
\IEEEpubid{\begin{minipage}{\textwidth}\ \\[12pt]
  FERMILAB-CONF-20-426-SCD
\end{minipage}} 
\title{FPGAs-as-a-Service Toolkit (FaaST)}

\author{\IEEEauthorblockN{Dylan Rankin, Jeffrey Krupa, \\
Philip Harris}
\IEEEauthorblockA{\textit{Massachusetts Institute of Technology}\\
Cambridge, MA 02139, USA}
\and
\IEEEauthorblockN{Maria Acosta Flechas, Burt Holzman,  \\
Thomas Klijnsma, Kevin Pedro, \\
Nhan Tran}
\IEEEauthorblockA{\textit{Fermi National Accelerator Laboratory}\\
Batavia, IL 60510, USA}
\and
\IEEEauthorblockN{Scott Hauck, Shih-Chieh Hsu, \\
Matthew Trahms, Kelvin Lin, Yu Lou}
\IEEEauthorblockA{\textit{University of Washington}\\
Seattle, WA 98195, USA}
\and
\IEEEauthorblockN{Ta-Wei Ho}
\IEEEauthorblockA{\textit{National Tsing Hua University}\\
Hsinchu, Taiwan 300044, R.O.C.}
\and
\IEEEauthorblockN{Javier Duarte}
\IEEEauthorblockA{\textit{University of California San Diego}\\
La Jolla, CA 92093, USA}
\and
\IEEEauthorblockN{Mia Liu}
\IEEEauthorblockA{\textit{Purdue University}\\
West Lafayette, IN 47907, USA}
}

\maketitle
\begin{abstract}
Computing needs for high energy physics are already intensive and are expected to increase drastically in the coming years.
In this context, heterogeneous computing, specifically as-a-service computing, has the potential for significant gains over traditional computing models.
Although previous studies and packages in the field of heterogeneous computing have focused on GPUs as accelerators, FPGAs are an extremely promising option as well.
A series of workflows are developed to establish the performance capabilities of FPGAs as a service.
Multiple different devices and a range of algorithms for use in high energy physics are studied.
For a small, dense network, the throughput can be improved by an order of magnitude with respect to GPUs as a service.
For large convolutional networks, the throughput is found to be comparable to GPUs as a service.
This work represents the first open-source FPGAs-as-a-service toolkit.
 
\end{abstract}


\input{intro}
\input{relatedwork}
\input{inference}
\input{results}

\input{discuss}
\input{conclusions}

\appendix
\input{adaeappendix.tex}

\section*{Acknowledgements}

We acknowledge the Fast Machine Learning collective as an open community of multi-domain experts and collaborators. 
This community was important for the development of this project. 
We would like to thank Steven Timm for his support of our work with HEPCloud.

M.~A.~F., B.~H., T.~K., K.~P., and N.~T. are supported by Fermi Research Alliance, LLC under Contract No. DE-AC02-07CH11359 with the U.S. Department of Energy (DOE), Office of Science, Office of High Energy Physics.
N.~T. is partially supported by the DOE Early Career Award.  
K.~P. is partially supported by the High Velocity Artificial Intelligence grant as part of the DOE High Energy Physics Computational HEP sessions program.
P.~H., and D.~R. are supported by NSF grants \#1934700,  \#1931469, and the IRIS-HEP grant \#1836650. 
J.~K. is supported by NSF grant \#190444. 
Cloud credits for this study were provided by Internet2 managed Exploring Cloud to accelerate Science (NSF grant \#190444). 
S.-C. H. is supported by the DOE Early Career Award under the grant DE-SC0015971. 
K.~L. is supported by NSF grants \#1934360.
Cloud credits for this study were provided by Internet2 managed Exploring Cloud to accelerate Science (NSF grant \#190444) and the Department of Energy Early Career Award. 
J.~D. is supported by the DOE Office of Science, Office of High Energy Physics Early Career Research program under Award No. DE-SC0021187. 

\bibliographystyle{IEEEtran}
\bibliography{fpgaaas}

\end{document}

%% file: intro.tex
\section{Introduction}
\label{sec:intro}
The breakdown of Dennard scaling~\cite{dennard} in the last decade has changed the landscape of modern computing~\cite{breakdown}.
Without the promise of ever-faster central processing units (CPUs) at a fixed power consumption, users have been forced to search elsewhere for solutions to their ever-growing computing needs~\cite{octwiki, octwiki_atlas}.
Some improvements in processor performance have come from the advent of multi-core processors.
However, there is growing interest in alternative computing architectures, such as graphics processing units (GPUs), field-programmable gate arrays (FPGAs), and application-specific integrated circuits (ASICs).
All of these architectures have been used in the past for various specialized tasks that make explicit use of their specific strengths, but a broader range of use cases has been encouraged in recent years by heterogeneous computing.
Heterogeneous computing denotes systems which make use of more than one type of computing architecture, typically a CPU and one of the alternative architectures above.
The alternative architecture is typically referred to as the ``coprocessor'' or ``accelerator.''
The advantage of this computing paradigm is that each algorithm can be run on the best-suited architecture.
Tools and strategies to simplify the use of heterogeneous computing solutions have enabled a growing list of applications to take advantage of alternative architectures.

Implementations of heterogeneous computing can take various forms.
The simplest design is typically to connect each accelerator to a CPU, and then have each CPU offload some portion of its work to the accelerator.
However, this is not necessarily the most effective design for a given system. 
One alternative paradigm, called computing ``as a service'', consists of separate server and client CPUs~\cite{EaaS,cloud_aas,service}.
Server CPUs are directly connected to the accelerators, and are responsible only for managing requests to communicate with the accelerator.
Client CPUs have network connections to the servers and are responsible for all other parts of the computing workflow; to use the accelerator they must send requests and receive replies from the servers.
This design separates the management of the accelerator from rest of the workflow, and simplifies the integration of the accelerator.
Replacing the accelerated application with a request to and reply from the server allows the client to remain insensitive to specifics of the accelerator such as the architecture, physical connections, transfer protocols, and other details of handling the data.

Heterogeneous workflows involving GPUs have been used for machine learning (ML) with great success~\cite{10.1145/2788396}, but workflows involving FPGAs have been slower to develop.
Traditionally, algorithm development for FPGAs has been restricted to experts well-versed in hardware description languages (HDLs), greatly limiting the pool of possible developers.
Conversely, high-level synthesis (HLS) compilers are capable of transforming untimed C into applications written in HDL, reducing the barrier to entry for FPGA algorithm development~\cite{hls_survey}.
For certain tasks, modern HLS tools are able to achieve performance comparable to that of handwritten HDL~\cite{Ghanathe_2017}.

Despite their relative immaturity as accelerators in heterogeneous workflows, FPGAs have many appealing features from a computing standpoint.
Fast algorithms can be run in nanoseconds, allowing large speedups in comparison to the same algorithms on CPUs.
FPGAs are also capable of running many smaller operations in parallel, thus allowing further improvements in speed.
Although ASICs are capable of providing similar or better factors of improvement in terms of speed, the ability to customize FPGAs allows them to be adapted to many different tasks or updated as algorithms and needs change.
FPGAs are also capable of providing this performance with reduced power consumption when compared to CPUs or GPUs.

As with GPU-based heterogeneous computing tools, many tools focused on FPGAs are designed with ML algorithms in mind, specifically deep neural networks (DNNs).
The characteristics of most ML algorithms, specifically a small number of inputs and a large number of operations, are well-suited for as-a-service computing models.
The algorithms considered in our work are all ML algorithms of different sizes, meant to span a wide range of possible requirements and design parameters.
All of the algorithms explored here are contenders for integration into heterogeneous workflows involving FPGAs.

We use a combination of custom and existing tools intended to target the specific needs of each algorithm.
These are packaged into a cohesive set of implementations that contain both the server and client code required to deploy both small and large DNN models, with different NN architectures, on multiple different hardware platforms.
We refer to this as the \textit{FPGAs-as-a-Service Toolkit} (\faast)~\cite{faast_facile,faast_resnet}. 
The framework we employ and the server design are capable of supporting both ML and non-ML algorithms.

The rest of this paper is structured as follows. 
In Section~\ref{sec:related}, we review related work.
Section~\ref{sec:descripton} describes the set of tools and ML models used.
In Section~\ref{sec:results}, we give results for the \faast approach and compare it to other approaches with GPUs and CPUs.
Finally, Sections~\ref{sec:discuss}~and~\ref{sec:outlook} provide discussion and outlook.

%% file: relatedwork.tex
\section{Related Work}
\label{sec:related}
As-a-service computing for ML algorithms is a growing area of development at the intersection of the fields of ML and on-demand cloud computing~\cite{cloud,servicemanifesto}.
The bulk of the tools available focus mainly on accelerating inference for large convolutional neural networks (CNNs) using GPUs.
Our work builds directly on some of these existing platforms.

High energy physics (HEP) workflows typically process an event using distinct modules, each responsible for executing a specific algorithm or computing a particular property of the event.
These modules can depend on the output of other modules, and therefore must be scheduled and in some cases run in a particular order to process an event successfully~\cite{IntelTBB}.
The Services for Optimized Network Inference on Coprocessors (SONIC)~\cite{SonicSW} approach is designed with HEP workflows in mind.
With this approach, the client API for a given server is integrated into an experiment's C++-based software framework, specifically the Compact Muon Solenoid (CMS) experiment at the CERN Large Hadron Collider.
Notably, SONIC supports accelerating generic algorithms using asynchronous, non-blocking methods.
This allows event processing on the CPU to proceed simultaneous with the accelerated algorithm, making maximal use of the computing resources.

The feasibility of the as-a-service computing model for HEP workflows has been previously demonstrated using SONIC to interact with a GPU-based server for inference~\cite{Krupa:2020bwg}.
The server/client design employed within this paper is similar to previous work, allowing for a direct comparison of the performance.
In addition, the design similarity showcases the versatility of the SONIC framework to handle both GPU and FPGA-based coprocessor servers.

Our results utilize multiple DNNs that are all relevant for HEP.
These networks span a range of sizes, use cases, and constraints.
For small networks, the server is implemented using \hlsfml~\cite{Duarte:2018ite,vladimir_loncar_2020_3969548} and Vitis Accel (previously SDAccel)~\cite{Kathail2020}.
For large networks, the server is implemented using Xilinx ML Suite~\cite{XilinxML}.

The \hlsfml package translates neural network models into FPGA firmware.
The firmware description is generated in an HLS language and is then compiled into a firmware description in VHDL/Verilog.
\hlsfml contains various tunable parameters to control the resource usage and performance, which are very useful to maximize the performance of the design.
Vitis Accel is a tool designed by Xilinx to allow for implementation and acceleration of generic FPGA kernels and their management from a host CPU.
In some of our work, the server and its communication with the FPGA is written using Vitis Accel, while the FPGA kernels to perform the inference are created with \hlsfml.

Xilinx ML Suite~\cite{XilinxML} is a library developed by Xilinx that can deploy CNNs to Xilinx FPGAs. 
It contains a utility to quantize models, a compiler that coverts \textsc{TensorFlow}~\cite{TensorFlow} or \textsc{Caffe}~\cite{caffe} models to an internal format, a CNN processing unit implementation on FPGA, and a \python interface.

Azure Machine Learning~\cite{azurewhitepaper} is a cloud-based environment developed by Microsoft to train, deploy, and manage ML models.
It provides, among other things, a \python software development kit to interact with the Microsoft Azure Stack Edge (ASE)~\cite{msdbedatasheet}, a physical on-premises network appliance capable of providing several ML models as a service.

%% file: inference.tex
\section{Tool Description}
\label{sec:descripton}

\begin{figure*}[t!]
\centering
\includegraphics[width=\textwidth]{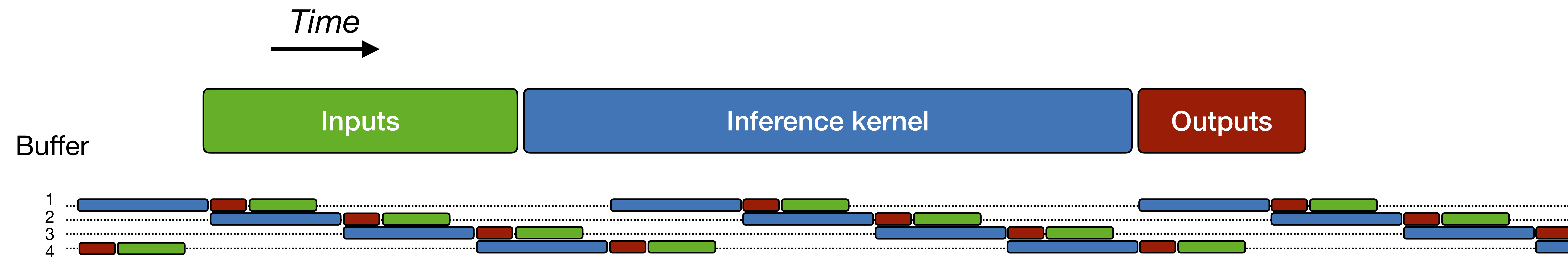}
\caption{Schematic of the task schedule for a DDR buffer size equal to 4 times a single batched input. 
The scheduling is shown after the buffers have stabilized after startup.}
\label{fig:facile_ddr_buffer}
\end{figure*}

We use the SONIC framework to implement the client.
The client employs asynchronous non-blocking \grpc calls to send requests to the server~\cite{gRPC}.

In order to perform inference on FPGAs, we use a combination of commercial and self-developed tools.
We design services for two benchmark networks: FACILE and ResNet-50.
These networks differ dramatically in size and design constraints, and therefore we use separate methodologies to construct a service for each.
For FACILE, we use \hlsfml and Vitis Accel, while for ResNet-50, we use either Xilinx ML Suite or Azure Machine Learning Studio.

For both networks, we first use the same formatting for client-server messages and requests as the Nvidia Triton Inference Server~\cite{Triton}.
This allows the exact same client to be used with either GPUs or FPGAs as a service with no modifications.
In the case of ResNet-50, we also investigate an alternative server design (still using \grpc) that runs on the Microsoft Azure Stack Edge.

In many cases, we find that the server performance is limited first by the server itself.
Specifically, even without performing any acceleration or explicit processing in the server, the throughput is limited by the \grpc server's ability to accept requests and return replies.
In order to remove this limitation, we employ proxy servers to allow multiple server instances to run simultaneously while remaining visible to the client as a single entity.

\begin{figure*}[t!]
\centering
\includegraphics[width=\textwidth]{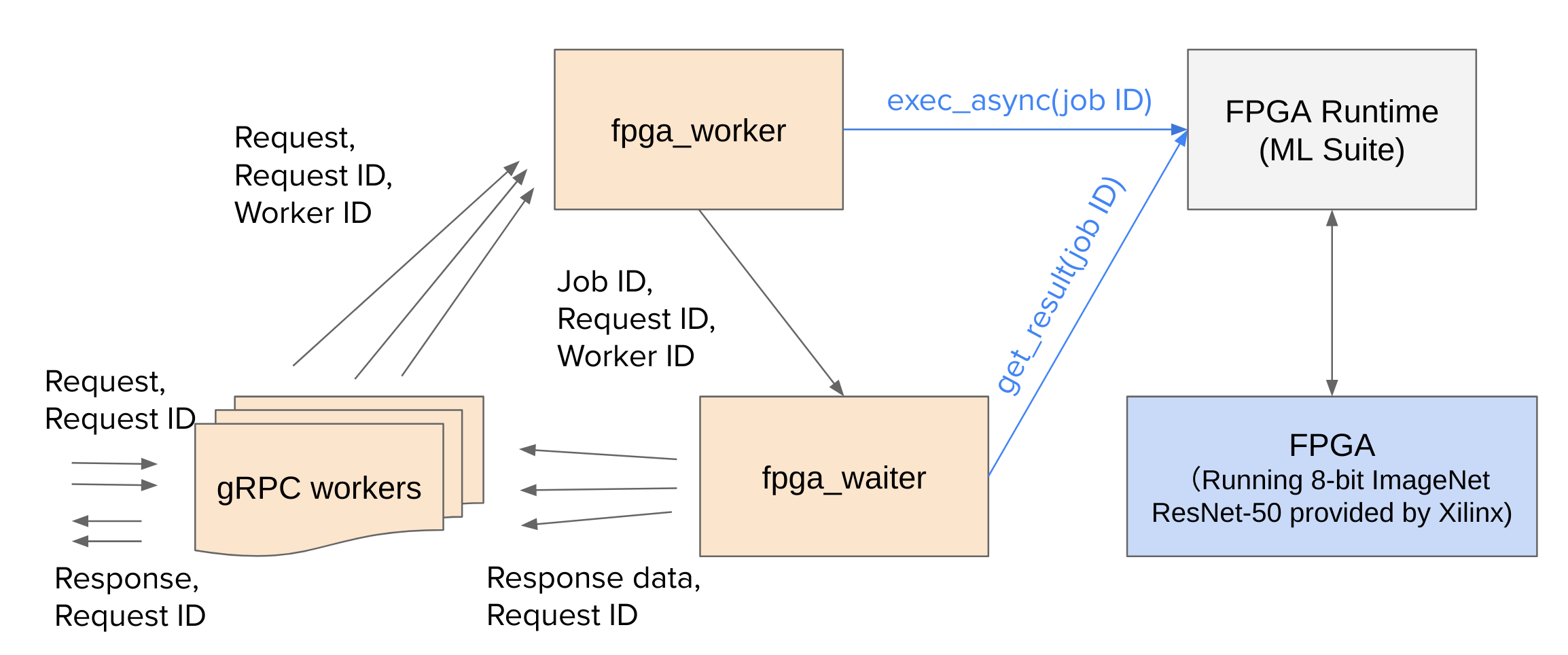}
\caption{Server structure. \grpc workers communicates with the Internet. The FPGA worker sends the data to the FPGA. The FPGA waiter waits for the job completion signal.}
\label{fig:ml-suite-server-structure}
\end{figure*}

\subsection{FACILE}

FACILE is a small fully-connected neural network trained to regress the energy of a particle based on time-sequence data read out from a calorimeter, an experimental apparatus that measures the energy a particle loses as it passes through it~\cite{Fabjan:2003aq}.
The network takes multiple measurements of the energy deposited in a region of the CMS hadron calorimeter as input and outputs the incident particle's energy.
This type of regression task is very common, with many algorithms of various sizes and different architectures employed across HEP experiments~\cite{Rovere:2020rqi, ACAT2019}.
FACILE is quite compact, consisting of three hidden layers with widths 31, 11, and 3 and rectified linear unit (ReLU) activation functions~\cite{relu}, three batch normalization layers~\cite{batchnorm}, and 1,001 trainable parameters, and therefore serves as a useful benchmark for ultrafast accelerated algorithms.
The synthesized FPGA kernel accepts all inputs simultaneously and produces the output in 34 clock cycles, with a clock frequency of 300\unit{MHz}.
This means that the inference result is available in 104\unit{ns}.
This application can also be run with a batched input consisting of all 16,000 channels of the calorimeter.
In order to provide inputs to and receive outputs from the kernel running on an FPGA, we use Vitis Accel.

Vitis Accel provides a software framework to manage signals between an FPGA and a CPU.
The core of this framework is a ``shell'' on the FPGA that connects the programmable logic to external memories, such as double data rate synchronous dynamic random-access memory (DDR SDRAM) banks, and to the CPU via a PCI Express (PCIe) connection.
The necessity of implementing the inference kernel inside this framework places various restrictions on its design.
The shell's design requires that the FPGA kernels are connected to the CPU only through the external DDR memory banks.
As a result, the inputs and outputs are not sent directly from the CPU to the FPGA kernel, but instead are streamed from the CPU to the DDR SDRAM via PCIe, and then from the DDR SDRAM to FPGA kernel.
This design means that the kernel must be capable of buffering inputs and outputs until they are all present on the chip for a given inference.
Although some computations necessary for the inference result can proceed without the full set of inputs available, the overall latency will still be dictated by the last-arriving input.
Further, for small networks especially, the resources and performance cannot be improved meaningfully by adapting the kernel design specifically for streaming inputs.

Vitis Accel also provides a framework for managing the data transfers (between the CPU and DDR memory) and the inference execution.
This is performed through the use of execution queues, where the dependence of a given task on previous tasks can be fully specified.
This means that the queue may contain tasks that are blocking or non-blocking; the call to place a task in the queue itself is non-blocking.
The three main tasks to queue are, in order, memory migration of the inputs from the CPU to the FPGA DDR SDRAM, kernel execution, and memory migration of the outputs from the FPGA DDR SDRAM to the CPU.
The simplest command flow to execute successfully is to make each of these three calls blocking.
The result is that they execute sequentially, and a new inference may only begin once the result of the previous inference call is received.

There are two main improvements that can be made to this basic command flow.
The first is to fully utilize the FPGA resources.
The two chips used in this work are the Xilinx Virtex UltraScale+ VU9P FPGA via Amazon Web Services (AWS) Elastic Compute Cloud (EC2) F1 instances, and the Xilinx Alveo U250 Data Center Accelerator Card.
Both of these chips are large modern FPGAs constructed from multiple super logic regions (SLRs) with limited connections between SLRs.
As a result, it is significantly simpler to design algorithms that can be placed on a single SLR.
The VU9P comprises 3 SLRs, while the Alveo U250 comprises 4 SLRs.
Since the \hlsfml kernel above can be placed on a single SLR, it is straightforward to place 3 (4) copies of the kernel on a VU9P (Alveo U250).
The copies of the kernel are referred to as ``compute units'' (CUs) in the language of Vitis Accel.
In order to avoid the need for crossing SLR boundaries to access the DDR memory, we must also restrict each CU to access only the DDR memory connected directly to the SLR on which it is placed.
By creating multiple CUs, this design can provide a proportional improvement in the throughput that can be achieved.

The other improvement that can be made is to make more effective use of the task queue.
This can be done by using a buffer in each DDR memory bank.
Since the total time for the kernel execution with large batch exceeds the total time for memory migration from the CPU to the FPGA DDR SDRAM, we define a region in the DDR memory with size equal to an integer multiple of the size of a single input batch.
Then for each CU, instead of requiring that the three main tasks are executed sequentially, we copy inputs such that the DDR buffer is always full.
This allows each CU to execute continuously by iterating sequentially through the DDR buffer.
Some tracking of the CU completion is still necessary for the memory migration of outputs from the FPGA DDR SDRAM to the CPU as well as at startup when the DDR buffer is not yet full.
Figure~\ref{fig:facile_ddr_buffer} shows the schedule for a buffer size of four inputs once the design is running stably.
The optimal scheduling for this design requires that the input buffer always contains an input when the inference kernel is available.
If the ratio of the total transfer time to the kernel execution time is $R$, we expect that the buffer must at least be large enough for $R$ inputs to ensure that optimal scheduling is possible.

Finally, we find that the \grpc server itself cannot handle more than approximately 2,000 requests per second.
In order to increase this limit, we spawn 8 threads to handle inference requests, each listening on a distinct address but sharing the same task queue for the FPGA.
We then use a HAProxy server~\cite{haproxy} to accept requests on a single address and forward requests in a round-robin fashion to the 8 addresses that correspond to the threads above.
This configuration allows the \faast server to fully utilize the FPGA.


\subsection{ResNet-50}

ResNets belong to a class of neural network architectures that use the residual learning technique~\cite{resnet}, with ResNet-50 denoting a particular version with 50 layers. 
While ResNet-50 was initially designed for natural image classification, it has been adapted to many other types of problems.
In this work, we use a ResNet-50 model trained to classify collimated showers of particles, or jets, generated from proton-proton collisions~\cite{kasieczka_gregor_2019_2603256,Duarte:2019fta}.
Specifically, the model is trained to distinguish jets originating from a top quark from other jets.
Similar image-based algorithms have been shown to be very effective at this particular classification task~\cite{Kasieczka_2019}.
Large networks like ResNet-50 are a useful benchmark in contrast to FACILE, since they require much longer latencies and therefore represent a different class of possible as-a-service use cases.
To construct the image used as input, we map the detector's surface to a two-dimensional grid and assign each pixel's value to be the total transverse momentum detected at the corresponding position. 
For this task, after the primary ResNet-50
feature extractor resulting in 2,048 features, a custom classifier is added, which
comprises one fully connected layer of width 1,024
with ReLU activation and another fully connected
layer of width 2 with softmax activation, whose output represents the probability of the jet arising from a top quark or not.

\subsubsection{Xilinx ML Suite}

To provide ResNet-50 as a service, we first used Xilinx ML Suite to quantize and load the model. 
We considered Vitis AI, but, at the time of writing, Xilinx did not officially support Vitis AI on AWS. 
Although we did not convert our ResNet-50 model for top quark tagging to the format used by Xilinx ML Suite, the default ResNet-50 model has a similar number of parameters and operations. 
Therefore, we expect that the performance in terms of latency and throughput should be similar.

ML Suite is used with an asynchronous inference call. Each inference request to the FPGA is assigned a job ID to identify the request. 
We restrict the server so that at most 8 jobs are in process simultaneously; other requests are queued. 
To utilize the asynchronous feature, we create two threads that communicate with ML Suite. 
The first thread, called the FPGA worker, fetches new data from a pending job's queue and passes it to the ML Suite runtime as soon as there is an available job ID. Another thread, called the FPGA waiter, waits for a job's completion signal and then fetches the inference result when it becomes available.
Figure~\ref{fig:ml-suite-server-structure} shows the workflow inside the server process.


As with FACILE, the public \grpc interface is the same as the Nvidia Triton server. 
Thus, existing SONIC clients can connect to this server without any modification.

\subsubsection{Azure Stack Edge}

A second method to provide ResNet-50 as a service is tested via an ASE.
Its main accelerator component is an Intel Arria 10 FPGA, to which several ML models may be deployed via the Azure Machine Learning Studio. 
No HLS or HDL is necessary, at the cost of not being able to run arbitrary ML models.
Additionally it includes a dual-core CPU, 12\unit{TB} storage, four 10/25\unit{GbE} network interfaces, and 128\unit{GB} of RAM.
The ASE was installed in the Feynman Computing Center at Fermi National Accelerator Laboratory and connected to the local network with a 10\unit{GbE} connection.
The ASE has a builtin network interface that accepts requests using the \grpc protocol.
Inference requests were sent to the ASE using the \grpc client implemented in SONIC.
To reduce any effects the networking might have on the latency, we exclusively used locally-connected CPU nodes for any inference requests.
We deployed the quantized version of the ResNet-50 top quark tagging model as provided in the Azure Machine Learning Studio software.

%% file: results.tex
\section{Results}
\label{sec:results}
\subsection{FACILE}

\begin{figure}[htbp]
\centering
\begin{subfigure}{\columnwidth}
\includegraphics[width=\columnwidth]{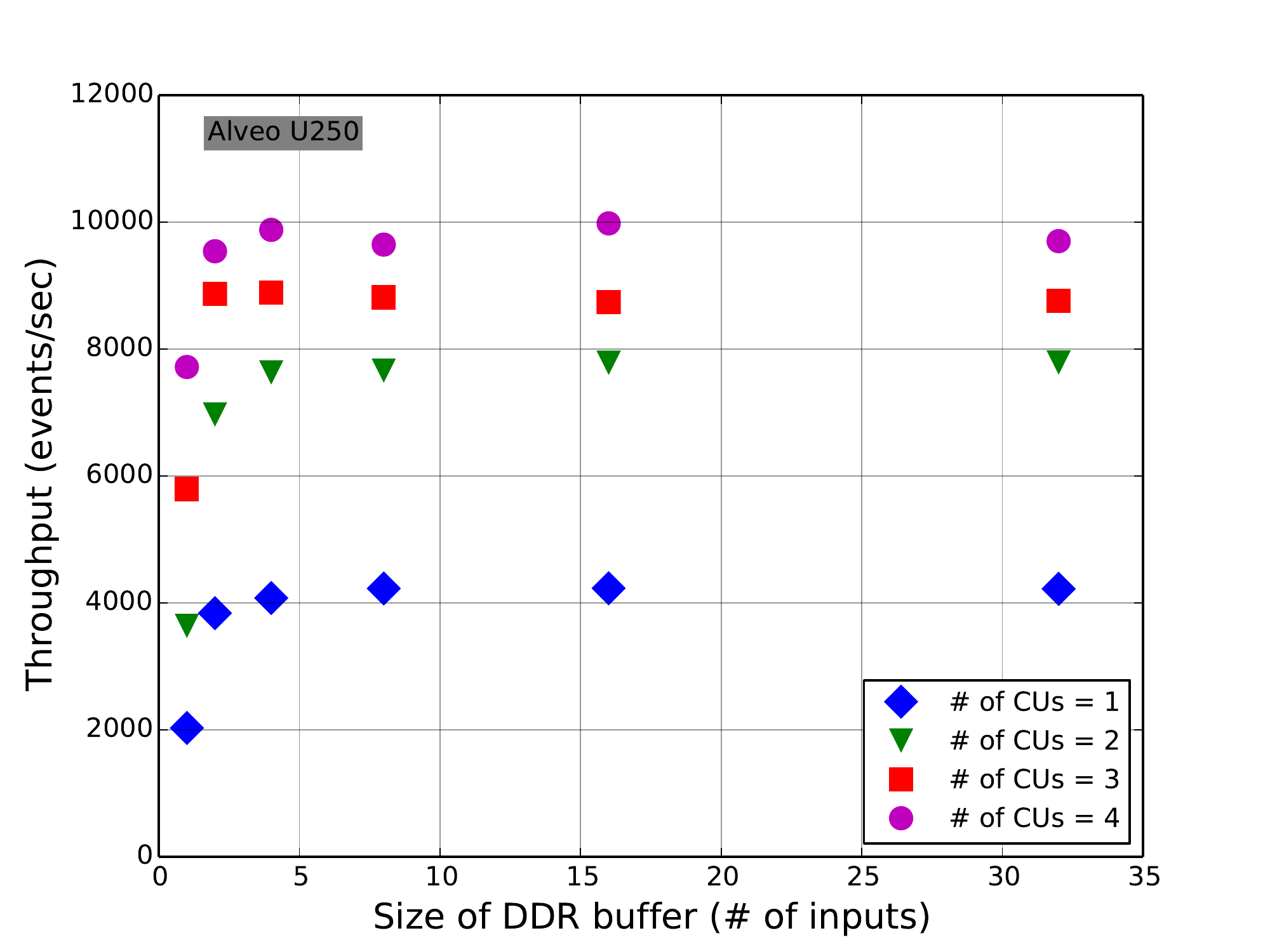}
\caption{Alveo}
\label{fig:alveo_scan_trpt}
\end{subfigure}
\begin{subfigure}{\columnwidth}
\includegraphics[width=\columnwidth]{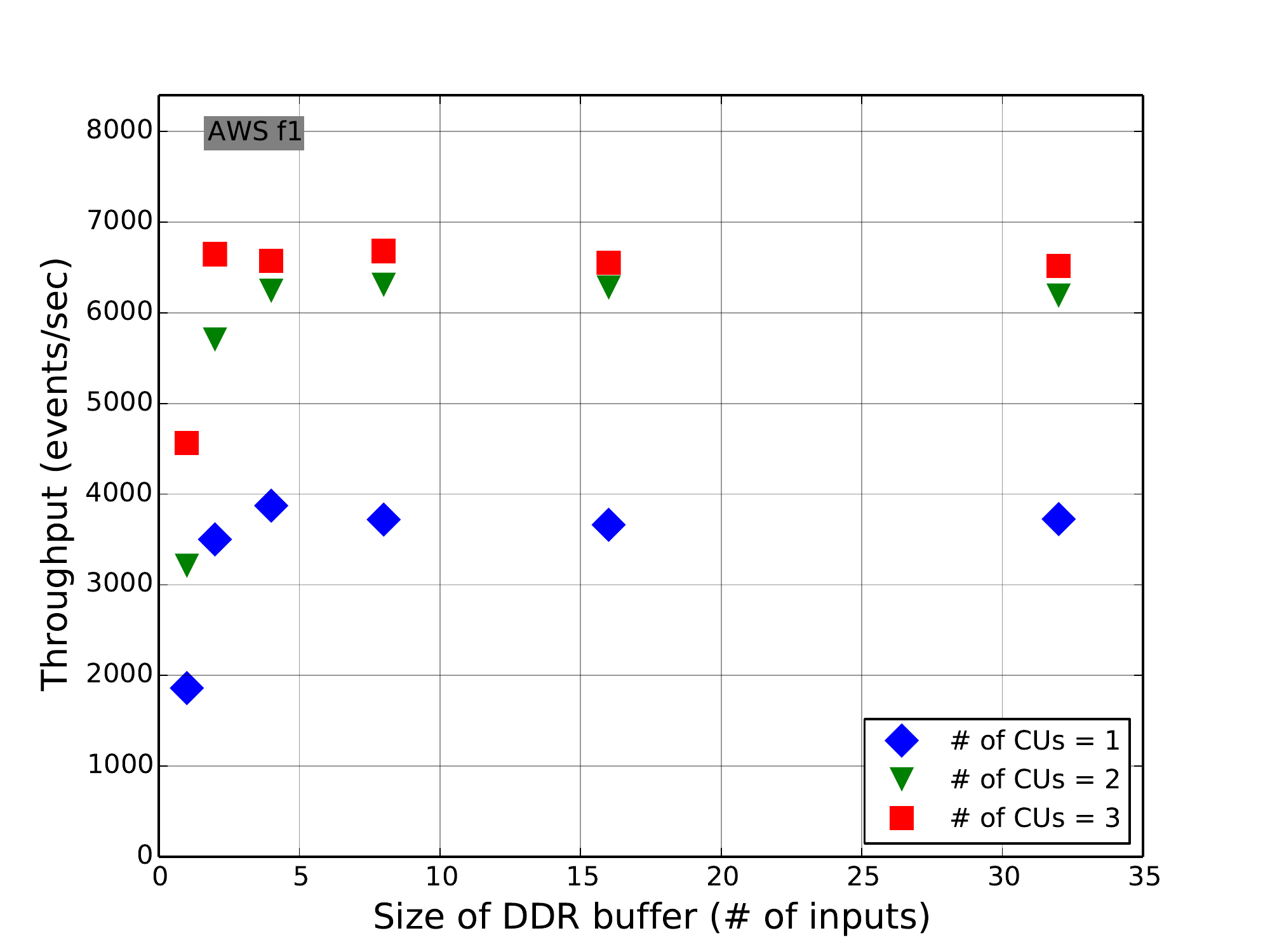}
\caption{F1 instance}
\label{fig:f1_scan_trpt}
\end{subfigure}
\caption{Throughput achieved locally for different numbers of CUs and sizes of the DDR buffer for the Alveo (a) and AWS EC2 F1 instance (b).}
\label{fig:scan_trpt}
\end{figure}

In order to evaluate the maximal theoretical throughput for a \faast server running FACILE, we built a custom application combining the server and client.
The values of the inputs for the test were determined during initialization and left unchanged throughout the test.
This design removes both the transfer and input preprocessing steps, and ensures that the throughput is limited only by FPGA inference capabilities.
We then use this application to scan a range of values for the size of the DDR buffer and number of CUs.
The results are shown in Fig.~\ref{fig:scan_trpt} for both an Alveo U250 and an AWS EC2 F1 instance.
We confirm that using more CUs allows higher throughput, and observe that throughput is maximal for DDR buffer sizes larger than 4 input batches.
This is expected because the ratio of the total transfer time to the kernel execution time for this design is roughly 3.
We also confirm the expectation that using buffer sizes larger than this optimal size does not affect the server performance.
These settings motivate our ultimate \faast server design that uses a DDR buffer size of 8 inputs (in units of the batch) and one CU per SLR on the device (3 for the AWS EC2 F1, 4 for the Alveo U250).
With these settings we are able to achieve a throughput of approximately 10,000 events per second using the Alveo U250 and 6,700 events per second using the AWS EC2 F1.

\begin{figure}[htbp]
\centering
\includegraphics[width=\columnwidth]{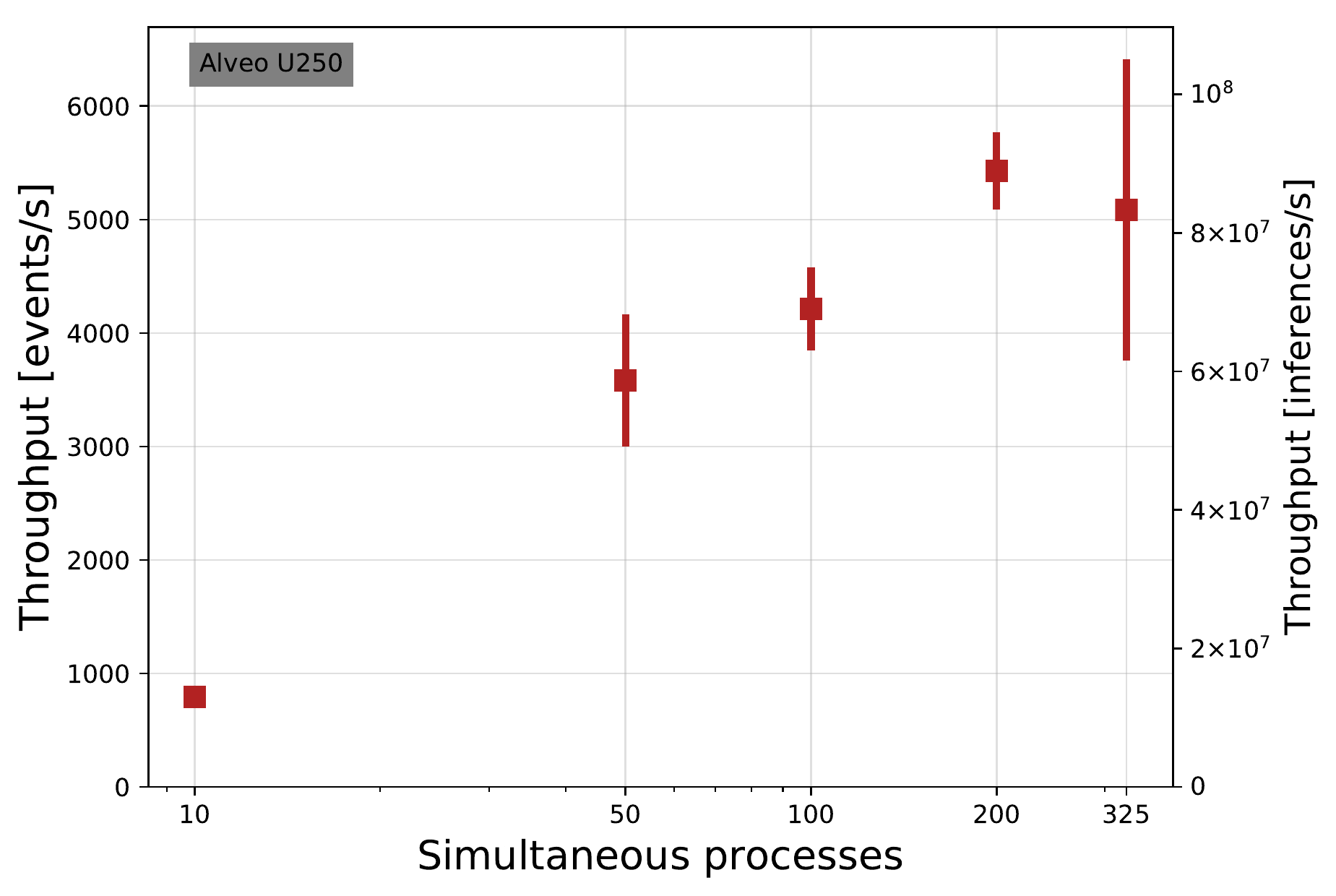}
\caption{Throughput achieved by a \faast server with a Alveo U250 for different numbers of simultaneous clients.}
\label{fig:alveo_test1_trpt}
\end{figure}

We use these settings to perform two tests of the \faast server performance.
For the first test, we run a workflow involving only the SONIC client module, and use a \faast server running on an Alveo U250.
The clients and server are both located at Fermi National Accelerator Laboratory.
The performance of the server is measured for varying numbers of simultaneous clients and the results are shown in Fig.~\ref{fig:alveo_test1_trpt}.
We find that the server is capable of running at a throughput of over 5,000 events per second, or 80 million inferences per second.
This is significantly below the maximal throughput possible for the Alveo U250 alone.
Despite the optimizations included in the server design, we find that the server CPU still limits the overall throughput.
This is largely a consequence of the small size (low latency) and large batch for FACILE; we expect that for most algorithms the CPU should be able to process requests fast enough to saturate the FPGA kernel.

This first test is useful for understanding the maximum throughput possible with a \faast server running FACILE since the workflow involves no tasks that can be performed asynchronously to the accelerated module.
However, this is not representative of most workflows, in which there are many tasks that either cannot be accelerated or are simply better performed on the CPU.
In this case, the CPU is able to schedule other tasks while the server processes the requests, thereby masking some of the latency of the acceleration.
Therefore, the second test examines the feasibility of using \faast in a realistic HEP setting, namely the CMS high-level trigger (HLT), which is the second tier of the trigger system implemented in software running currently entirely on CPUs.  
It is responsible for performing a reconstruction of the full detector, but this must be done quickly, with latency on the order of 100\unit{ms}. 
This is therefore a good candidate for usage with \faast.
This second test is completed by running the full CMS HLT workflow and bechmarking the default HLT configuration to one with the hadron calorimeter reconstruction performed using the SONIC client and \faast server described above.
For reproducibility in the HLT tests, we use the HEPCloud framework which allows various experiments to run analysis workloads on demand in the public cloud as well as some allocation-based high-performance computing (HPC) sites~\cite{Holzman2017}.
We deploy the HLT client jobs in the form of AWS EC2 \texttt{r4.4xlarge} instances. 
These client Virtual Machines are provisioned with 16 High Frequency Intel Xeon E5-2686 v4 (Broadwell) processors, 122\unit{GiB} DDR4 Memory and support for enhanced networking. 

\begin{figure}[htbp]
\centering
\includegraphics[width=\columnwidth]{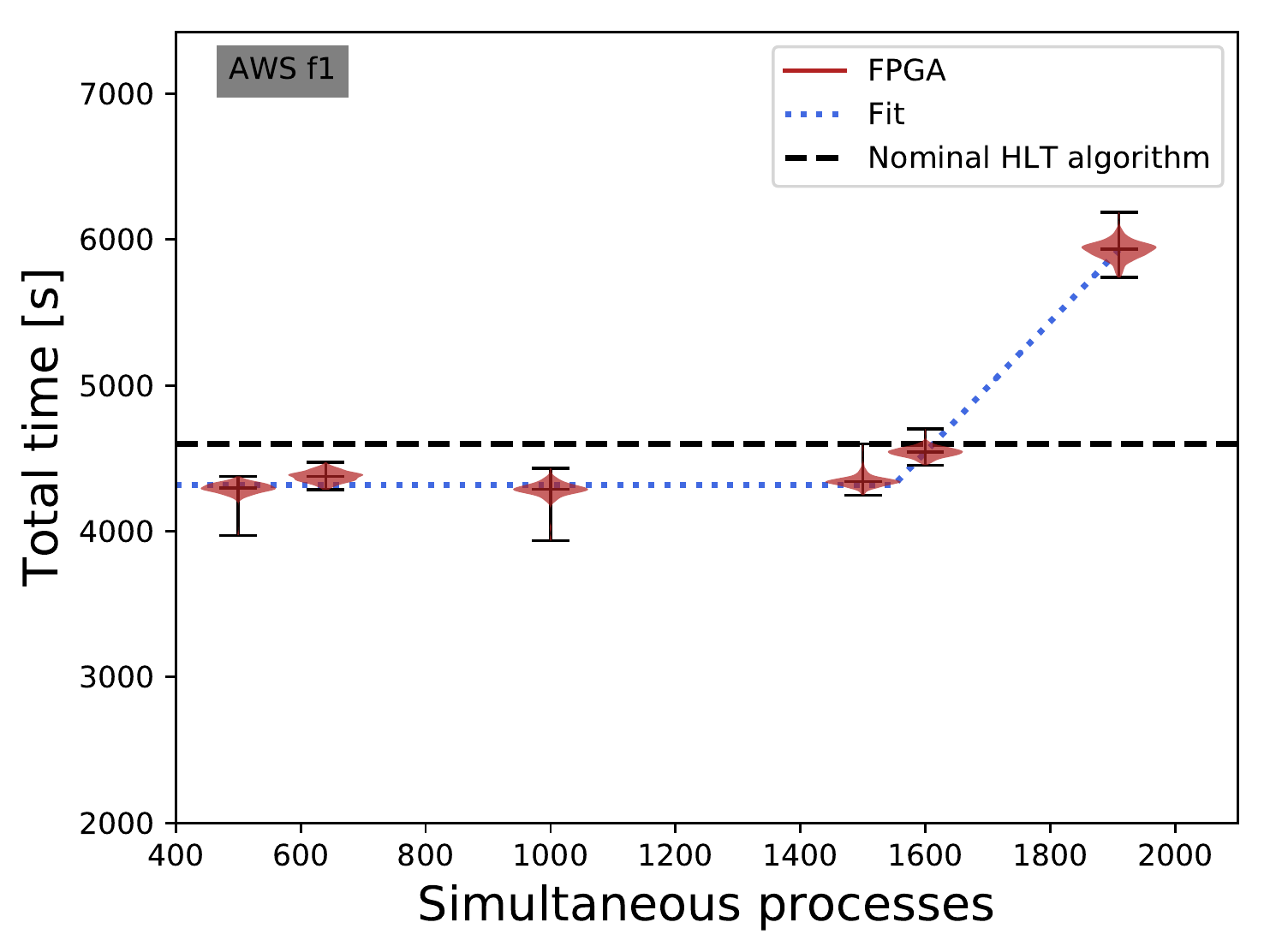}
\caption{Total processing time required for running a realistic HLT workflow using the FACILE \faast server as a function of the number of simultaneous clients. The black dashed line represents the total processing time required for running the HLT workflow with no as-a-service component. The blue dotted line displays a piecewise linear fit to the measurements.}
\label{fig:facile_hlt_test}
\end{figure}

We find that a single \faast server running FACILE is capable of serving 1500 simultaneous clients without any increase in processing time.
Above 1500 simultaneous clients, we find that the 25\unit{Gbps} network bandwidth limit of an AWS \texttt{f1.16xlarge} introduces delays in processing from the as-a-service model.
Based on the results achieved using an Alveo U250 with a 100\unit{Gbps} network bandwidth limit, we estimate that one \faast server could serve approximately 3,600 simultaneous HLT processes with no reduction in performance.


\subsection{ResNet-50}

\subsubsection{Xilinx ML Suite}

To maximize the throughput achievable with Xilinx ML Suite, we investigated an alternative to the nominal \grpc request interface called ``StreamInfer.''
In this type of connection, the server can receive a stream of images and returns a stream of inference results. 
We found that this type of connection is more efficient than the standard ``Infer'' requests because it avoids the overhead of reconnecting the server for every request. 
On an AWS \flarge instance, the streaming connection's throughput is 17\% higher (487 inference/second using 8 FPGAs) compared to the standard connections when multiple clients are connected.

\begin{figure}[t]
    \centering
    \includegraphics[width=\columnwidth]{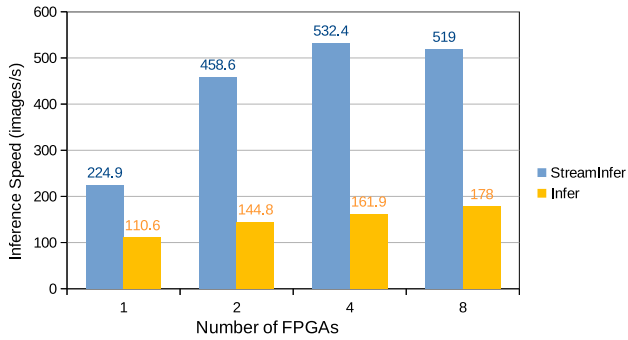}
    \caption{FPGA scaling test using a single ML Suite runtime instance and a single client. ``SteamInfer'' denotes a streaming connection, while ``Infer'' indicates a standard connection. The server is run on an AWS \flarge instance.}
    \label{fig:fpga-scaling-single-instance}
\end{figure}

Although the Xilinx ML Suite runtime supports connecting to multiple FPGAs, we found that the server performance does not increase proportionally with the number of FPGAs.
We found no significant performance gain when connecting to more than 2 FPGAs simultaneously.
Figure~\ref{fig:fpga-scaling-single-instance} shows results with ``StreamInfer'' and ``Infer'' requests when connecting to various numbers of FPGAs.


We suspect that this poor scalability is caused by the \python Global Interpreter Lock (GIL), which limits the server to use only one CPU core at a time.
To bypass this limit, we started 4 server processes on the same machine, each connecting to only 2 FPGA cards.
An Nginx load balancer, also running on the same machine, is then used to distribute the requests to each process~\cite{nginx}.
Since each inference request is much larger than that of common use cases for \grpc, we need to increase Nginx's buffer size to achieve optimal performance.

\begin{figure}[t]
\centering
    \includegraphics[width=\columnwidth]{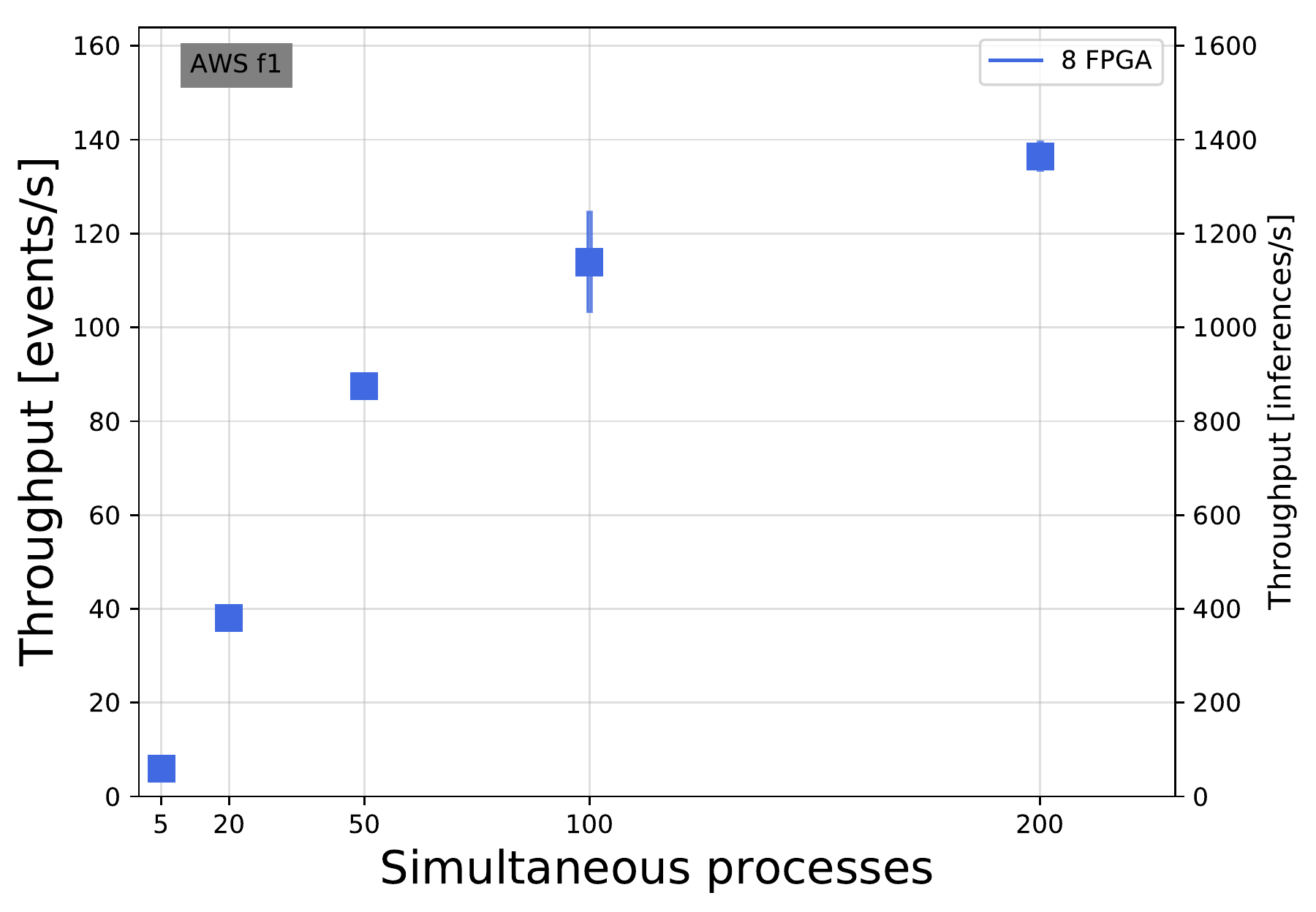}
    \caption{FPGA scaling test result. Simultaneous processes means the number of client instances running at the same time. Each event contains a batch of 10 images.}
    \label{fig:ml-suite-scaling-test-nginx}
\end{figure}

To verify our design's scalability, we ran a server on an AWS \flarge instance in the \texttt{us-west} region.
This type of instance is connected to 8 FPGAs.
We used a cluster at Fermi National Accelerator Laboratory to issue requests to the server from multiple instances of SONIC clients.
Results from these tests are shown in Fig.~\ref{fig:ml-suite-scaling-test-nginx}.
Our design is able to achieve a 550\% improvement in throughput when using 8 FPGAs (1,350 inferences/second) compared to a single FPGA (220 inferences/second).


\subsubsection{Azure Stack Edge}

\begin{figure}[t]
\centering
\includegraphics[width=\columnwidth]{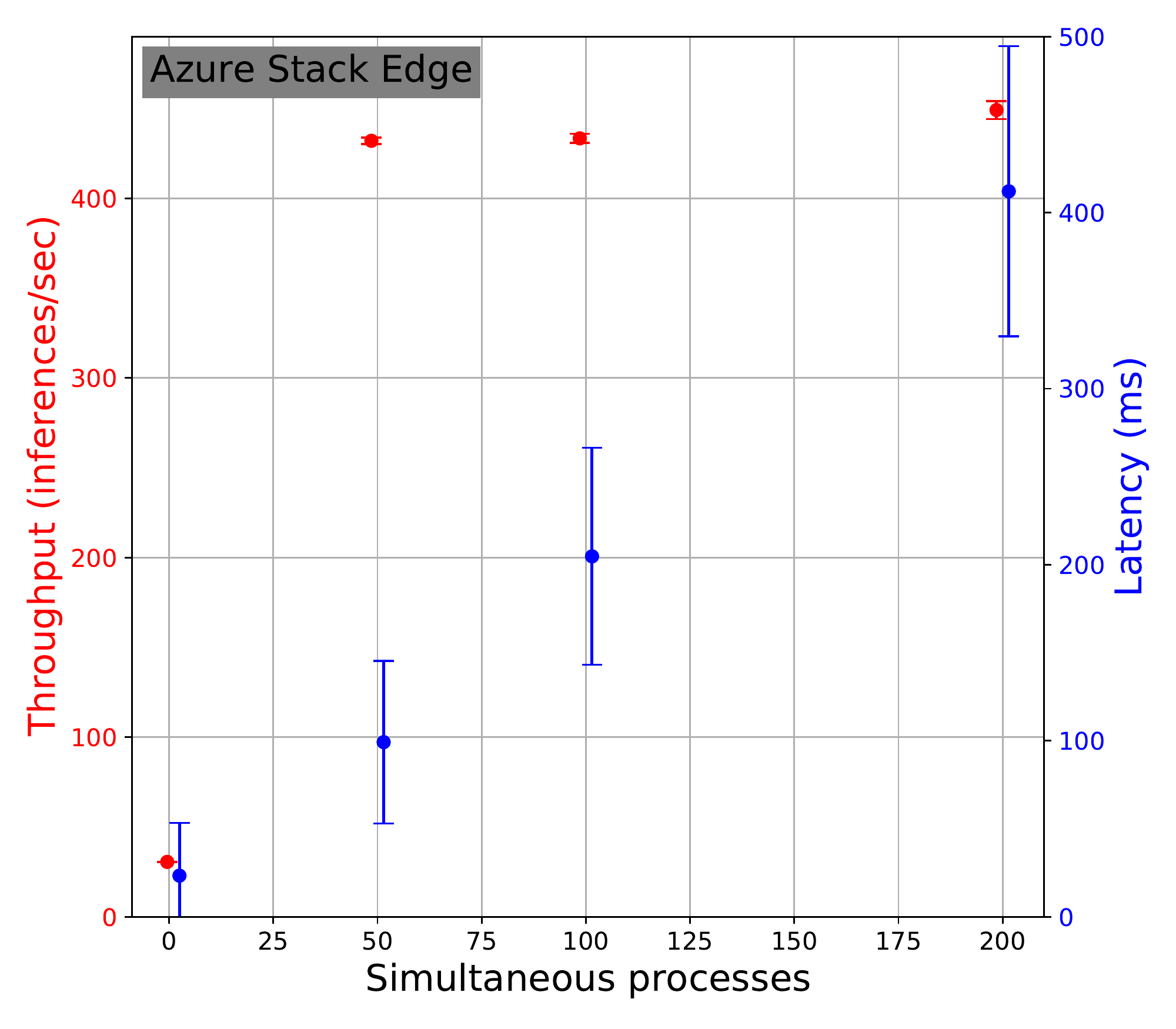}
\caption{
Throughput (red, left axis) and latency (blue, right axis) as a function of number of simultaneous processes sending inference requests of the ResNet-50 model to the Azure Stack Edge.
}
\label{fig:ase}
\end{figure}

In order to measure the achievable throughput of the ASE providing the ResNet-50 model as a service, we used 200 CPUs concurrently sending inference requests via the local network at Fermi National Accelerator Laboratory.
We find that, using SONIC to send the inference requests of our benchmark ResNet-50 model, the average throughput of the ASE is $449.2 \pm 5.0$ inferences/second, with a maximum achieved throughput of 460.1 inferences/second.
Using less than 200 cores reduces the throughput slightly: with 50 cores, the average throughput is $432.0\pm 1.7$ inferences/second.
As expected for a fully utilized FPGA, the latency, measured to be the time difference between the start of the inference request and the time a response is received, depends approximately linearly on the number of simultaneous processes sending inference requests.
For 50 (200) cores, we find an average latency of $99.2 \pm 46.2\unit{ms}$ ($412.1 \pm 82.4\unit{ms}$).
Sending requests with a single CPU severely underutilizes the FPGA, but yields a picture of the minimum achievable latency.
We find a mean latency of $23.4 \pm 30.0\unit{ms}$ when using a single core, noting that the latency is not normally distributed but actually strongly influenced by networking effects.
In a minority of inference requests, the latency jumped to 100\unit{ms} or larger, which is solely attributable to network effects, and unrelated to the inference time of the ASE.
The median of the distribution, which is less affected by these high-latency outliers, is 12.7\unit{ms}.
The throughput and latency as a function of number of simultaneous processes are shown in Fig.~\ref{fig:ase}.

Finally, the ASE's own CPU can be used as a client to perform the inference on the internal FPGA.
By sending inference requests from the internal CPU to the internal FPGA of the ASE, we find an average throughput of 70 inferences per second, or 14\unit{ms} per inference.
It should be noted that we used the Azure Machine Learning \python SDK rather than SONIC for this test, as it was technically less complex to deploy on the ASE CPU.
The throughput in this test is largely driven by the extent to which the CPU manages to utilize the full FPGA.
It is nevertheless a helpful comparison for the large-scale test described above.

%% file: discuss.tex
\section{Discussion}
\label{sec:discuss}

We present the FPGAs-as-a-service toolkit (\faast) for integrating FPGA-based machine learning (ML) inference as a service into scientific workflows.
We have shown examples of how \faast can be used for a broad range of applications and hardware.
A summary of the results for all implementations are shown in Table~\ref{tab:summary}.
For large networks, we find that the throughput of \faast servers is comparable to or better than similar GPU as-a-service designs.
In the case of small dense networks, such as FACILE, a \faast server outperforms GPU as-a-service implementations by over an order of magnitude.
These results are not contingent on the precise details of the networks we use as benchmarks.
Indeed, we expect similar performance from \faast for other network inference applications.
\faast represents the first open source toolkit intended to make high performance FPGAs as-a-service available generically.

\begin{table}[htbp]
\caption{Summary of the performance of \faast servers in terms of events and inferences per second, and bandwidth. Results for performance on GPUs are taken from Ref.~\cite{Krupa:2020bwg}.}
\centering
    \resizebox{\columnwidth}{!}{
\begin{tabular}{llc|ccc}
\hline\noalign{\smallskip}
\multirow{2}{*}{Algorithm} & \multirow{2}{*}{Platform}& Number of & Batch & Inf./s & Bandwidth \\
 & & Devices & Size & [Hz] & [Gbps] \\
\noalign{\smallskip}\hline\noalign{\smallskip}
FACILE  & AWS EC2 F1 & 1  & 16,000 & 36\,M & 23 \\
FACILE & Alveo U250 & 1  &  16,000 & 86\,M & 55 \\
FACILE  & T4 GPU & 1  &  16,000 & 8\,M & 5.1 \\
ResNet-50 & AWS EC2 F1 & 8  & 10 & 1400 & 6.7  \\
ResNet-50 & V100 GPU & 8 & 10 & 1,700 & 8.1  \\
ResNet-50 & ASE & 1 & 1 & 460  & 2.2  \\
ResNet-50 & T4 GPU & 1  & 10 & 250 & 1.2  \\
\noalign{\smallskip}
\hline
\end{tabular}
}
\label{tab:summary}
\end{table}

For inference on GPUs, performance gains with respect to CPUs typically occur in tasks that can be run with large batch sizes.
This is due to the ability of the GPU to run many parallel operations.
FPGAs, on the other hand, do not gain exclusively by using large batches.
Rather, FPGAs are able to achieve low inference latency as a result of their ability to perform computations significantly faster than CPUs and GPUs.
As a result, for ResNet-50, our \faast server running on the ASE with batch 1 almost doubles the throughput when compared to a T4 GPU running with batch 10.
This is especially noteworthy given that many tasks in high energy physics (HEP) workflows that require complex algorithms are naturally run with low batch size.
For example, in the case of the top quark tagging ResNet-50 model used in this work, a batch size of 2 may be sufficient for most HEP events.

One caveat to the performance of FPGAs with small batches is that transfers to and from the device are typically more efficient for large batches.
This is because the overhead for transfers can be quite significant.
For a similar network to FACILE, inference at batch size 1 was found to be only 15 times faster than inference at batch 16,000~\cite{aigean}.
However, not every ML algorithm should be run at maximum batch; this latency improvement must be weighed against the additional resources and infrastructure needed to handle a large number of concurrent inputs on the FPGA.

We have exclusively used ML applications in this work because of their widespread and growing use in HEP workflows, as well as their ability to be parallelized.
This makes them very useful target applications for acceleration.
However, the \faast server design is highly generic.
Provided that an algorithm can be successfully executed on an FPGA, the \faast model is capable of enabling as-a-service acceleration.
Any functional FPGA kernel can be accelerated using Vitis Accel in a similar manner to FACILE.

%% file: conclusions.tex
\section{Outlook}
\label{sec:outlook}

FPGAs have been traditionally been used for various specialized tasks.
Their low power consumption and extremely fast processing make them particularly suited for applications across industry and high energy physics.
Their advantages, however, are not exclusive to these domains and can be leveraged for many other high-performance computing tasks.
The FPGAs-as-a-service toolkit we present can assist in the implementation of FPGAs as a service in a variety of computing workflows across science.

%% file: adaeappendix.tex
\section{Artifact Description / Artifact Evaluation Appendix}

\subsection{Artifact Description}

We ran tests of the FACILE hardware kernel throughput at Fermi National Accelerator Laboratory (FNAL) on a Xilinx Alveo U250 running XRT 2.3.1301 and Vitis 2019.2, with the hardware installed locally to a Intel Xeon Silver 4210 CPU @ 2.20GHz running Scientific Linux release 7.8. 
Tests of the FaaST server for FACILE v1.0.0 were run using this same machine for the server and the batch submission nodes at the FNAL LHC Physics Center (LPC) Computing Cluster for the clients. 
Tests of FACILE in a realistic workflow were run using HEPCloud using an AWS f1.16xlarge instance for the server and \texttt{r4.4xlarge} instances for the clients. 
Tests of ResNet-50 in Xilinx ML Suite were run using our FaaST interface v0.5.0, with a \flarge instance for the FaaST server and the batch submission nodes at the FNAL LPC Computing Cluster for the clients. 
Tests of ResNet-50 with the Azure Stack Edge were run locally at the FNAL Feynman Computing Center, using the batch submission nodes at the FNAL LPC Computing Cluster for the clients.

Our author-created artifacts are given in Ref.~\cite{faast_facile} and Ref.~\cite{faast_resnet}.

\subsection{Artifact Evaluation}

In all cases we ensure that behavior in critical regions (i.e. high throughput) can be reproduced with slightly different test settings, thus verifying that the results are both stable and reliable.
We run using a large number of events for all tests to ensure the accuracy of and reduce statistical uncertainties on our results.
All our results are expected to be generalizable to other networks and applications with similar performance.
They are cross checked on multiple similar devices whenever possible to ensure the stability with respect to machine specifications or device conditions and details.
We use monitoring tools for cloud tests to ensure no significant issues are occurring that could affect our results. 
For tests run using FNAL resources we have good control of the machines and devices in use and can ensure that there are no transient sources impacting the results. 
We also run results over the course of hours or days such that any intermittent issues should not persist across data points.